\documentclass[cameraready]{Interspeech}
\usepackage{multirow}
\usepackage{textcomp}
\usepackage{mathtools}
\usepackage{amsmath}

\definecolor{lightgreen}{RGB}{204,255,204}
\usepackage[table]{xcolor}

\usepackage{siunitx}

\title{YingMusic-Singer: Controllable Singing Voice Synthesis with
Flexible Lyric Manipulation and Annotation-free Melody Guidance}

\author[affiliation={1,2}, orcid=0009-0005-5957-8936]{Chunbo}{Hao}
\author[affiliation={2}, orcid=0009-0003-2602-2910]{Junjie}{Zheng}
\author[affiliation={1}, orcid=0009-0001-6706-0572]{Guobin}{Ma}
\author[affiliation={1}]{Yuepeng}{Jiang}
\author[affiliation={1}]{Huakang}{Chen}
\author[affiliation={1}]{Wenjie}{Tian}
\author[affiliation={2}, orcid=0009-0003-9258-4006]{Gongyu}{Chen}
\author[affiliation={2}, orcid=0009-0005-5413-6725]{Zihao}{Chen}
\author[affiliation={1}, correspondingauthor]{Lei}{Xie}

\address{
  $^1$ Audio, Speech and Language Processing Group
  (ASLP@NPU)\\ School of Computer Science, Northwestern Polytechnical
  University, China \\
  $^2$ AI Lab, Giant Network, China
}

\email{cbhao@mail.nwpu.edu.cn, zhengjunjie@ztgame.com, lxie@nwpu.edu.cn}


\begin{document}

\maketitle

\begin{abstract}

  Regenerating singing voices with altered lyrics while preserving
  melody consistency remains challenging, as existing methods either
  offer limited controllability or require laborious manual
  alignment. We propose YingMusic-Singer, a fully diffusion-based
  model enabling melody-controllable singing voice synthesis with
  flexible lyric manipulation. The model takes three inputs: an
  optional timbre reference, a melody-providing singing clip, and
  modified lyrics, without manual alignment. Trained with curriculum
  learning and Group Relative Policy Optimization, YingMusic-Singer
  achieves stronger melody preservation and lyric adherence than
  Vevo2, the most comparable baseline supporting melody control
  without manual alignment. We also introduce LyricEditBench, the
  first benchmark for melody-preserving lyric modification
  evaluation. The code, weights, benchmark, and demos are publicly
  available at \url{https://github.com/ASLP-lab/YingMusic-Singer-Plus}.
\end{abstract}

\section{Introduction}

Singing Voice Synthesis (SVS) aims to generate human-like singing
voices from musical scores, lyrics, and timbre references. Modern
systems~\cite{DBLP:conf/aaai/Liu00CZ22, DBLP:conf/acl/HeLYHCLZ23,
  DBLP:conf/aaai/ZhangHLHXCDHZ24, DBLP:conf/iscslp/WangBDXGL24,
DBLP:conf/slt/YuSWTW24, DBLP:conf/acl/ZhangGPYZ0W0025} achieve
high-fidelity synthesis, yet most rely on precisely annotated paired
data associating each phoneme with an exact pitch contour and
duration. While such fine-grained control is indispensable for
professional music production, preparing these annotations creates a
prohibitive barrier for an increasingly important use case,
\emph{singing voice editing}, which regenerates an existing singing
voice with modified lyrics while preserving the original melodic and
rhythmic structure. This capability is highly desirable for song
adaptation, personalized cover generation, rapid prototyping of vocal
arrangements, and multilingual song localization.

Existing editing paradigms fall into two categories.
The first adopts an in-context learning strategy that masks the
region to be edited and regenerates it conditioned on the surrounding
context and target lyrics~\cite{lei2024songcreator,yang2025songeditor}.
Although convenient, this approach is restricted to local segments
and provides limited melody control. The second category, widely
adopted in practice, relies on commercial SVS models such as
Synthesizer {V}\footnote{\url{https://dreamtonics.com/synthesizerv}}
and {ACE} Studio\footnote{\url{https://acestudio.ai}}, where users
first need to transcribe the original vocal into MIDI, then assign
the modified lyrics to the notes and durations, and re-synthesize the
audio. This pipeline offers strong controllability but demands manual
effort, and the complexity increases for tasks such as cross-lingual
translation. In summary, existing approaches either depend on the
surrounding context to recover melody or require manual alignment of
word-level timestamps with melody information, limiting their
flexibility and scalability.
Several recent efforts address these challenges.
Vevo2~\cite{zhang2025vevo2unifiedcontrollableframework} achieves
melody-controllable generation, yet suffers from reduced
intelligibility and poor melody adherence.
SoulX-Singer~\cite{qian2026soulxsingerhighqualityzeroshotsinging}
supports existing singing as melody input but still requires manual
alignment of word-level timestamps, leaving the core burden
unresolved. While these recent efforts narrow the gap, they either
achieve suboptimal performance or still need manual alignment.

To address these challenges, we propose \emph{YingMusic-Singer}, a
fully diffusion-based model for melody-preserving lyric editing.
First, YingMusic-Singer introduces a streamlined editing paradigm
that synthesizes singing voices from only three inputs, namely an
optional timbre reference clip, a singing clip providing the target
melody, and the corresponding modified lyrics, without the need for
manual alignment or precise annotation. Second, to mitigate limited
phoneme generalization caused by the small scale and challenging
vocal techniques in singing data, and to address the inherent
trade-off between faithful lyric reproduction and melody adherence,
we employ a curriculum training strategy combined with Group Relative
Policy Optimization (GRPO)~\cite{DBLP:journals/corr/abs-2402-03300,
guo2025deepseek}, enabling strong performance on both dimensions
simultaneously. Third, we construct \emph{LyricEditBench} based on
GTSinger~\cite{DBLP:conf/nips/ZhangPGLZWXLHWZ24}, the first benchmark
for lyric modification evaluation under matched melody conditions,
covering six common editing scenarios with balanced sampling for fair
and comprehensive comparison.
Experiments show that YingMusic-Singer outperforms
Vevo2~\cite{zhang2025vevo2unifiedcontrollableframework}, currently a
comparable alignment-free alternative, in both melody preservation
and lyric adherence. We publicly release the model weights, inference
code, and LyricEditBench to facilitate further research and advance
the practical adoption of singing voice editing.

\section{Methodology}

\subsection{Architecture Overview}

\begin{figure*}[htb]
  \centering
  \includegraphics{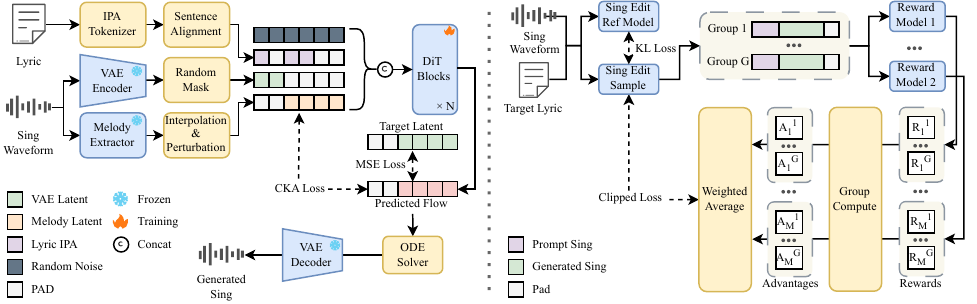}
  \caption{Overall architecture of YingMusic-Singer. Left: the
    training pipeline consisting of a Variational Autoencoder, a Melody
    Extractor, an IPA Tokenizer, and DiT-based conditional flow
  matching. Right: the GRPO training pipeline.}
  \label{fig:YingMusic-Singer}
  \vspace{-10pt}
\end{figure*}

As shown in Figure~\ref{fig:YingMusic-Singer}, YingMusic-Singer
generates singing voices at 44.1\,kHz from three inputs: an optional
timbre reference, a melody-providing singing clip, and corresponding
modified lyrics. It comprises: (1) a Variational Autoencoder (VAE)
following Stable Audio 2~\cite{DBLP:conf/ismir/EvansPCZTP24}, whose
encoder $\mathcal{E}$ downsamples a stereo 44.1\,kHz singing waveform
$\mathbf{x} \in \mathbb{R}^{T \times 2}$ by a factor of 2048 into
$\mathbf{z} = \mathcal{E}(\mathbf{x}) \in \mathbb{R}^{T' \times D}$,
and whose decoder $\mathcal{D}$ reconstructs high-fidelity audio
$\hat{\mathbf{x}} = \mathcal{D}(\mathbf{z})$ at inference; (2) a
Melody Extractor built upon the encoder of a pretrained MIDI
extraction model, whose intermediate representations naturally
capture disentangled melody information. It produces $\mathbf{h} =
\mathcal{M}(\mathbf{M}) \in \mathbb{R}^{L \times D_m}$, which is then
temporally interpolated to $\tilde{\mathbf{h}} \in \mathbb{R}^{T'
\times D_m}$ to match the VAE latent frame rate; (3) an IPA Tokenizer
that converts both Chinese and English lyrics into a unified discrete
phoneme sequence. To ensure the model correctly distinguishes between
prompt and generation regions, avoiding phoneme omission or
repetition at boundaries, we adopt sentence-level alignment following
DiffRhythm~\cite{ning2025diffrhythmblazinglyfastembarrassingly}. Each
lyric sentence is converted into an IPA subsequence and placed at its
corresponding onset frame within a padded frame-level sequence of
length $T'$. The aligned sequence is passed through a learnable
embedding layer to yield $\mathbf{e} \in \mathbb{R}^{T' \times D_e}$.
During inference, prompt lyrics are placed at the beginning of the
sequence and target lyrics at the start of the masked region,
requiring no timestamp annotation from the user; and (4) a
\textbf{DiT-based CFM backbone} following
F5-TTS~\cite{DBLP:conf/acl/ChenN0DWZ0025}.

During training, a proportion $\gamma$ of VAE latent frames is
randomly masked as the synthesis target, with the unmasked portion
serving as timbre context. Let $\mathbf{z}_{\text{ctx}}$ denote the
unmasked VAE latent (zero-filled in masked regions). The condition
$\mathbf{c} = [\tilde{\mathbf{h}};\, \mathbf{e};\,
\mathbf{z}_{\text{ctx}}]$ is concatenated with the noisy latent
$\mathbf{z}_t = (1-t)\mathbf{z}_0 + t\mathbf{z}_1$ along the channel
dimension and fed into the CFM, which learns a velocity field via:
\begin{equation} \label{eq:cfm_loss}
  \mathcal{L}_{\text{MSE}} = \mathbb{E}_{t, \mathbf{z}_0,
  \mathbf{z}_1, \mathbf{c}} \lVert v_\theta(\mathbf{z}_t, t,
  \mathbf{c}) - (\mathbf{z}_1 - \mathbf{z}_0) \rVert^2.
\end{equation}

\subsection{Curriculum Training}
To mitigate limited phoneme generalization caused by the small scale
and challenging vocal techniques in singing data, YingMusic-Singer
first undergoes \emph{TTS Pretraining} without melody conditioning.
The subsequent \emph{Singing Voice Supervised Fine-Tuning (SFT)}
stage has two phases. Phase~1 enables sentence-level alignment on
singing data, allowing the model to adapt to the singing domain.
Phase~2 activates melody conditioning and introduces a Centered
Kernel Alignment (CKA) loss to enforce melody adherence. Given the
predicted $v_\theta$ and melody $\tilde{\mathbf{h}}$, CKA measures
their alignment via Gram matrices $\mathbf{K} = v_\theta
v_\theta^\top$ and $\mathbf{L} = \tilde{\mathbf{h}}\tilde{\mathbf{h}}^\top$:
\begin{equation}
  \mathcal{L}_{\text{CKA}} = 1 - \frac{\lVert \mathbf{K}^\top
  \mathbf{L} \rVert_F^2}{\lVert \mathbf{K}^\top \mathbf{K} \rVert_F
  \lVert \mathbf{L}^\top \mathbf{L} \rVert_F},
\end{equation}
\noindent
and the total phase~2 loss is $\mathcal{L}_{\text{SFT}} =
\mathcal{L}_{\text{MSE}} + \lambda \mathcal{L}_{\text{CKA}}$.

\subsection{Group Relative Policy Optimization}

While curriculum training achieves high performance, SFT Phase~2
simultaneously degrades PER, exposing a persistent trade-off.
$\mathcal{L}_{\text{MSE}}$ targets holistic latent reconstruction and
cannot isolate specific deficiencies for targeted optimization, while
adjusting $\lambda$ in $\mathcal{L}_{\text{CKA}}$ only shifts the
balance without resolving the trade-off itself. Moreover, inevitable
noise in large-scale singing data, such as backing vocals and quality
artifacts, caps model performance at the dataset ceiling. To overcome
these limitations, reinforcement learning (RL) becomes essential.
PPO requires a value network that is expensive to train. Offline
methods such as DPO suffer from distribution shift, as pre-collected
preference data becomes stale when the policy improves.
GRPO~\cite{DBLP:journals/corr/abs-2402-03300, guo2025deepseek}
operates online and estimates baselines from within-group reward
statistics, eliminating the value network while remaining efficient and stable.

Following recent
efforts~\cite{DBLP:journals/corr/abs-2505-05470,DBLP:journals/corr/abs-2507-21802,wang2026flowsegrpotrainingflowmatching},
we convert the deterministic ODE trajectory into an SDE but restrict
stochastic steps to a bounded window, ensuring precise advantage
attribution to exploratory steps. To prevent collapse toward a single
reward dimension, we employ $M$ reward models jointly and compute the
advantage for each sample as
\begin{equation}
  A^i = \sum_{k=1}^{M} w_k \frac{R^i_k -
  \mathrm{mean}(\{R^j_k\})}{\mathrm{std}(\{R^j_k\})},
\end{equation}
and the final GRPO loss is
\begin{equation}
  \mathcal{L}_{\text{GRPO}}(\theta) =
  \frac{1}{G}\sum_{i=1}^{G}\frac{1}{|S|}\sum_{t \in
  S}\left(-\mathcal{L}_{\text{clip}} + \beta\, D_{\mathrm{KL}}\right),
\end{equation}
where $\mathcal{L}_{\text{clip}}$ is the clipped surrogate objective
over the current-to-old policy likelihood ratio, $D_{\mathrm{KL}}$
regularizes the current policy toward the reference, $G$ is the group
size, $S$ the set of SDE sampling steps, and $\beta$ the KL
regularization strength.

\begin{table}[t]
  \centering
  \caption{Task types for lyric modification in LyricEditBench.}
  \vspace{-0.2cm}
  \label{tab:task_types}
  \footnotesize
  \begin{tabular}{llp{0.4\columnwidth}}
    \toprule
    \textbf{Abbr.} & \textbf{Task Type} & \textbf{Description} \\
    \midrule
    PSub & Partial Substitution & Substitute part of the words \\
    FSub & Full Substitution & Completely rewrite the song \\
    Del & Deletion & Delete some words \\
    Ins & Insertion & Insert some words \\
    Trans & Translation & CN $\leftrightarrow$ EN translation \\
    Mix & Code-Mixing & Mixed-language lyrics \\
    \bottomrule
  \end{tabular}
  \vspace{-14pt}
\end{table}

\begin{table*}[!t]
  \centering
  \caption{Comparison with Baseline Model on LyricEditBench across
    Task Types in Table \ref{tab:task_types} and Languages. Metrics
    (M): P: PER, S: SIM, F: F0-CORR, V: VS are detailed in
  Section~\ref{subsubsec:Objective_Metrics}. Best results are in \textbf{bold}.}

  \label{tab:model_comparison}

  \fontsize{8pt}{9.6pt}\selectfont
  \setlength{\tabcolsep}{4.2pt}

  \begin{tabular}{clccccccccccccc}
    \toprule
    \multirow{2}{*}{\textbf{Task}} & \multirow{2}{*}{\textbf{Model}}
    & \multirow{2}{*}{\textbf{M}} &
    \multicolumn{6}{c}{\textbf{Chinese}} &
    \multicolumn{6}{c}{\textbf{English}} \\
    \cmidrule(lr){4-9} \cmidrule(lr){10-15}
    & & & \textbf{PSub} & \textbf{FSub} & \textbf{Del} & \textbf{Ins}
    & \textbf{Trans} & \textbf{Mix} & \textbf{PSub} & \textbf{FSub} &
    \textbf{Del} & \textbf{Ins} & \textbf{Trans} & \textbf{Mix} \\
    \toprule
    \multirow{8}{*}{\shortstack{Melody\\Control\\(cross-timbre)}}
    & \multirow{4}{*}{Vevo2~\cite{zhang2025vevo2unifiedcontrollableframework}}
    & P $\downarrow$ & 0.1378& 0.1462& 0.1545& 0.1872& 0.4409&
    0.4757& 0.3352& 0.3481& 0.3812& 0.3135& 0.8019& 0.5132\\
    & & S $\uparrow$ & 0.6462& 0.6550& 0.6457& 0.6551&
    \textbf{0.6115}& \textbf{0.6490}& \textbf{0.6357}&
    \textbf{0.6161}& \textbf{0.6277}& \textbf{0.6359}&
    \textbf{0.6237}& \textbf{0.6325}\\
    & & F $\uparrow$ & 0.8471& 0.8188& 0.8345& 0.8552& 0.7678&
    0.8526& 0.8794& 0.8409& 0.8924& 0.8888& 0.8776& 0.8927\\
    & & V $\uparrow$ & 1.3578& 1.3784& 1.3491& 1.1346& 1.3061&
    1.4208& 1.0340& 1.1217& 1.0476& 0.9610& 1.0281& 1.0925\\
    \cmidrule(lr){2-15}
    & \multirow{4}{*}{Ours}
    & P $\downarrow$ & \textbf{0.0192}& \textbf{0.0197}&
    \textbf{0.0458}& \textbf{0.0208}& \textbf{0.0881}&
    \textbf{0.1563}& \textbf{0.0685}& \textbf{0.0692}&
    \textbf{0.1053}& \textbf{0.0716}& \textbf{0.0413}& \textbf{0.2668}\\
    & & S $\uparrow$ & \textbf{0.6543}& \textbf{0.6561}&
    \textbf{0.6489}& \textbf{0.6552}& 0.5791& 0.6395& 0.6078&
    0.5914& 0.6001& 0.5889& 0.5982& 0.6076\\
    & & F $\uparrow$ & \textbf{0.9364}& \textbf{0.9428}&
    \textbf{0.9351}& \textbf{0.9381}& \textbf{0.9378}&
    \textbf{0.9352}& \textbf{0.9355}& \textbf{0.9279}&
    \textbf{0.9309}& \textbf{0.9315}& \textbf{0.9290}& \textbf{0.9389}\\
    & & V $\uparrow$ & \textbf{2.0779}& \textbf{2.1419}&
    \textbf{2.1219}& \textbf{1.9887}& \textbf{1.9372}&
    \textbf{2.1002}& \textbf{1.5054}& \textbf{1.5418}&
    \textbf{1.6081}& \textbf{1.4036}& \textbf{1.5769}& \textbf{1.5060}\\
    \toprule
    \multirow{8}{*}{\shortstack{Sing Edit\\(self-timbre)}}
    & \multirow{4}{*}{Vevo2~\cite{zhang2025vevo2unifiedcontrollableframework}}
    & P $\downarrow$ & 0.1290& 0.1303& 0.1596& 0.1810& 0.4111&
    0.4659& 0.3414& 0.3538& 0.3531& 0.2944& 0.7680& 0.4951\\
    & & S $\uparrow$ & \textbf{0.7875}& \textbf{0.7729}&
    \textbf{0.8269}& \textbf{0.8324}& \textbf{0.7252}&
    \textbf{0.8015}& \textbf{0.7971}& \textbf{0.7729}&
    \textbf{0.8183}& \textbf{0.8378}& \textbf{0.7563}& \textbf{0.8346}\\
    & & F $\uparrow$ & 0.8858& 0.8805& 0.8969& 0.9115& 0.8456&
    0.9023& 0.9258& 0.9278& 0.9365& 0.9415& 0.9137& 0.9465\\
    & & V $\uparrow$ & 1.4860& 1.5100& 1.5094& 1.2935& 1.3535&
    1.4377& 1.0910& 1.1110& 1.1682& 1.1156& 1.1178& 1.1453\\
    \cmidrule(lr){2-15}
    & \multirow{4}{*}{Ours}
    & P $\downarrow$ & \textbf{0.0214}& \textbf{0.0186}&
    \textbf{0.0946}& \textbf{0.0426}& \textbf{0.1009}&
    \textbf{0.1903}& \textbf{0.0906}& \textbf{0.0782}&
    \textbf{0.1700}& \textbf{0.1070}& \textbf{0.0538}& \textbf{0.2946}\\
    & & S $\uparrow$ & 0.7622& 0.7392& 0.7874& 0.8028& 0.6539&
    0.7564& 0.7398& 0.7105& 0.7764& 0.7714& 0.6918& 0.7708\\
    & & F $\uparrow$ & \textbf{0.9615}& \textbf{0.9587}&
    \textbf{0.9628}& \textbf{0.9642}& \textbf{0.9542}&
    \textbf{0.9607}& \textbf{0.9610}& \textbf{0.9563}&
    \textbf{0.9675}& \textbf{0.9660}& \textbf{0.9498}& \textbf{0.9668}\\
    & & V $\uparrow$ & \textbf{1.9761}& \textbf{2.0345}&
    \textbf{1.8837}& \textbf{1.7689}& \textbf{1.9371}&
    \textbf{1.9283}& \textbf{1.4448}& \textbf{1.4086}&
    \textbf{1.3820}& \textbf{1.2553}& \textbf{1.4788}& \textbf{1.3464}\\
    \bottomrule
  \end{tabular}

  \vspace{-5pt}
\end{table*}

\subsection{LyricEditBench}

We build LyricEditBench from
GTSinger~\cite{DBLP:conf/nips/ZhangPGLZWXLHWZ24} by removing all
Paired Speech Group content, deduplicating audio via MD5 hashing, and
excluding clips exceeding 15 seconds. DeepSeek
V3.2~\cite{DBLP:journals/corr/abs-2512-02556} then generates modified
lyrics for each of the six modification types in
Table~\ref{tab:task_types}. Given original lyrics and modification
instructions, the LLM produces revised versions, with non-compliant
outputs discarded, yielding 11,535 valid samples. Samples are
classified by singer gender (male/female) and language
(Chinese/English) into four categories, then organized by
modification type. For each combination, we select 30 samples per
singing technique (covering the six techniques in GTSinger) and 120
for the technique-free category, resulting in 300 per modification
type per category and 7,200 test instances in total. For each
instance, a timbre prompt of no more than 15 seconds is randomly
drawn from the remaining audio pool, so each LyricEditBench instance
comprises a melody reference clip, a timbre prompt, and the
corresponding modified lyrics.

\section{Experimental Setup}
\textbf{Dataset.}

The Chinese and English subsets of
Emilia~\cite{DBLP:conf/slt/HeSWLGHLYLSWCZW24} are used for TTS
pretraining. For Singing Voice SFT, internally licensed music tracks
are processed by
SongFormer~\cite{hao2025songformerscalingmusicstructure} to segment
structural boundaries and label function categories, discarding
non-vocal segments. Vocal stems are then isolated using Mel-band
RoFormer~\cite{DBLP:journals/corr/abs-2310-01809}. We retain clips
between 2 and 30 seconds, splitting longer ones at sentence
boundaries, ultimately obtaining 33,562.6 hours of singing data. The
GRPO dataset is constructed by filtering SFT data with three
criteria: ASR transcript verification, retaining only clips with a
word error rate below 5\%, speaker diarization via
pyannote~\cite{DBLP:conf/interspeech/PlaquetB23,
DBLP:conf/interspeech/Bredin23}, keeping only single-speaker clips,
and a DNSMOS P.808 quality score~\cite{reddy2021dnsmos,
reddy2022dnsmos} threshold of 3.5. This yields approximately 20,240
curated clips with balanced Chinese and English content. The test set
in the proposed LyricEditBench is strictly excluded from training.

\textbf{Evaluation Metrics.}
\label{subsubsec:Objective_Metrics}
We evaluate models with four objective and two subjective metrics.
For objective evaluation, \emph{Phoneme Error Rate (PER)} measures
phoneme-level intelligibility, as singing exhibits lower semantic
density, fewer contextual cues, and greater pronunciation variation
than speech. Both Chinese and English clips are transcribed by
singing-trained Qwen3-ASR-1.7B~\cite{shi2026qwen3asrtechnicalreport}
and converted to phoneme sequences with tone markers removed.
\emph{Speaker Similarity (SIM)} follows
F5-TTS~\cite{DBLP:conf/acl/ChenN0DWZ0025}, extracting speaker
embeddings with a WavLM-large-based verification model and computing
cosine similarity. \emph{F0 Pearson Correlation (F0-CORR)} measures
melody adherence via frame-wise Pearson correlation between F0
contours of generated and reference clips using
RMVPE~\cite{DBLP:conf/interspeech/WeiCDC23}. \emph{Vocal Score (VS)}
adopts VocalVerse2~\cite{DBLP:conf/mm/WangYGLQLCFDZ25} as a learned
metric aligned with human perceptual preferences. For subjective
evaluation, 120 samples uniformly sampled across task types and
languages are rated by 30 listeners on two dimensions:
\emph{Naturalness Mean Opinion Score (N-MOS)} for overall perceptual
quality and naturalness, and \emph{Melody Mean Opinion Score (M-MOS)}
for faithfulness to the reference melody, both on a 5-point scale.

\textbf{Implementation Details.}
We adopt the VAE from Stable Audio
2~\cite{DBLP:conf/ismir/EvansPCZTP24} ($D=64$), the encoder of
SOME\footnote{\url{https://github.com/openvpi/SOME}} as Melody
Extractor ($D_m=128$, temporal dropout 0.1), and a DiT backbone
following F5-TTS~\cite{DBLP:conf/acl/ChenN0DWZ0025} (22 layers, 16
heads, hidden dim 1024, $D_e=512$). The full system has $\sim$727.3M
parameters (453.6M CFM, 156.1M VAE, 117.6M Melody Extractor), trained
on 8$\times$A800 80GB GPUs with DDP and bf16. Across all stages,
70\%--100\% of latent frames are randomly masked. TTS pretraining
runs for 1M steps (batch duration 1.268h, lr 1e-4). Singing Voice SFT
Phase~1 disables melody conditioning for 240K steps; Phase~2 enables
it for 170K steps ($\lambda$ decayed from 0.3 to 0.01 over the first
2K steps; batch duration 1.69h, lr 2.5e-5). For GRPO, $G{=}8$
candidates are scored by $M{=}4$ equally weighted reward models (SDE
  noise $a{=}0.8$, window $\mathcal{W}$ with $w_{\min}{=}1$, $w_s{=}8$,
$\epsilon_u{=}0.01$, $\epsilon_l{=}0.002$, $\beta{=}1$), optimized
for 1.2K steps (batch size 6, lr 7e-6) without CFG. Inference uses 32
ODE steps with CFG scale 3.

\section{Experimental Results}

\subsection{Main Results}

We compare against
Vevo2~\cite{zhang2025vevo2unifiedcontrollableframework}, a
token-based autoregressive model with disentangled timbre and melody
control. For Sing Edit (self-timbre), the timbre and melody
references share the same clip; for Melody Control (cross-timbre),
the timbre reference is replaced accordingly. Vevo2 is the most
direct baseline, as it supports the same functionalities under
comparable input conditions. Other recent systems operate under
fundamentally different paradigms. In-context learning approaches
require surrounding audio context with manually aligned edit
boundaries, restricting them to local segment editing. SoulX-Singer
relies on precise character-level timestamps; obtaining such
annotations via ASR introduces non-trivial character-count
mismatches, while ground-truth timestamps from curated datasets would
grant an unrealistic advantage absent in practice.

As shown in Table~\ref{tab:model_comparison}, YingMusic-Singer
consistently outperforms Vevo2 across all six modification types
under both Melody Control and Sing Edit settings in PER, F0-CORR, and
VS, demonstrating strong adherence to both the reference melody and
modified lyrics. The intelligibility gap is most pronounced on Trans
and Mix tasks, indicating that reconstructing a substantially
different phoneme sequence while preserving melody is inherently
challenging. Note that PER on Mix tasks may be further inflated by
ASR hallucinations on code-switched utterances. Vevo2's incomplete
melody disentanglement tends to reduce intelligibility and introduce
hallucinations, whereas YingMusic-Singer's unified IPA tokenization
and GRPO-based lyric adherence optimization maintain robustness even
under these extreme conditions. F0-CORR further distinguishes the two
systems: benefiting from CKA alignment and GRPO, YingMusic-Singer
maintains consistently high correlation across all tasks and
languages, whereas Vevo2 fluctuates considerably, suggesting less
robust melody control. For SIM, Vevo2 benefits from its multi-stage
architecture, where an autoregressive LLM handles melody and content
generation while a dedicated CFM focuses on timbre reconstruction,
effectively easing speaker modeling. YingMusic-Singer instead adopts
a single-stage CFM that jointly models all factors, prioritizing
architectural simplicity for practical deployment. However, the
higher SIM of Vevo2 comes at the cost of degraded PER and F0-CORR. In
practice, faithfully rendering modified lyrics while preserving the
original melodic structure remains the primary concern in lyric editing.
As shown in Table~\ref{tab:subjective}, YingMusic-Singer consistently
achieves higher N-MOS and M-MOS than Vevo2 across both tasks and
languages. The strong subjective scores also indicate that GRPO
optimization does not overfit to the reward models but generalizes to
human perception. Vevo2 receives lower ratings overall with notably
higher variance, and listeners report perceptible artifacts such as
unfaithful lyric rendering and melodic misalignment in its outputs,
suggesting less robust generation quality.

Beyond model comparison, the breadth of evaluation across six editing
types, two languages, and both objective and subjective metrics
further validates LyricEditBench as a comprehensive benchmark for
melody-preserving lyric editing, supporting future research on song
adaptation, cover generation, and cross-lingual vocal arrangement.

\begin{table}[t]
  \centering
  \caption{Subjective evaluation on LyricEditBench. N-MOS and M-MOS
  denote naturalness and melody adherence, respectively.}
  \fontsize{8pt}{9.6pt}\selectfont
  \setlength{\tabcolsep}{2pt}
  \label{tab:subjective}
  \begin{tabular}{ll cccc}
    \toprule
    \multirow{2}{*}{\textbf{Task}} & \multirow{2}{*}{\textbf{Model}}
    & \multicolumn{2}{c}{ZH} & \multicolumn{2}{c}{EN} \\
    \cmidrule(lr){3-4} \cmidrule(lr){5-6}
    & & N$\uparrow$ & M$\uparrow$ & N$\uparrow$ & M$\uparrow$ \\
    \midrule
    \multirow{2}{*}{\shortstack{Melody\\Control}}
    & Vevo2~\cite{zhang2025vevo2unifiedcontrollableframework} &
    4.25$\pm$0.06& 4.28$\pm$0.05& 4.31$\pm$0.05& 4.31$\pm$0.04\\
    & Ours & \textbf{4.31$\pm$0.04}& \textbf{4.44$\pm$0.04}&
    \textbf{4.36$\pm$0.05}& \textbf{4.51$\pm$0.03}\\
    \midrule
    \multirow{2}{*}{\shortstack{Sing\\Edit}}
    & Vevo2~\cite{zhang2025vevo2unifiedcontrollableframework} &
    4.48$\pm$0.05& 4.41$\pm$0.05& 4.44$\pm$0.04& 4.50$\pm$0.04\\
    & Ours & \textbf{4.52$\pm$0.04}& \textbf{4.55$\pm$0.04}&
    \textbf{4.55$\pm$0.04}& \textbf{4.58$\pm$0.03}\\
    \bottomrule
  \end{tabular}

\end{table}

\vspace{-0.3cm}
\subsection{Ablation Study}

As shown in Table \ref{tab:ablation}, the curriculum learning
pipeline introduces clearly separable improvements at each stage. TTS
Pretrain establishes articulatory priors but lacks singing capability
(F0-CORR near zero), with PER degrading significantly when a singing
clip serves as the ICL prompt, indicating poor domain adaptation. SFT
Phase~1 substantially improves all metrics, achieving the lowest PER
as the model adapts to the singing domain while freely generating
melody from the ICL prompt, bypassing the demanding task of explicit
melody alignment. F0-CORR under Sing Edit improves slightly, showing
partial melody capture from context alone, yet explicit guidance
remains necessary for faithful reproduction. SFT Phase~2 activates
the Melody Extractor and raises F0-CORR above 0.92, though PER
increases, reflecting the difficulty of jointly maintaining melody
fidelity and lyric faithfulness. GRPO resolves this trade-off by
recovering PER while further improving F0-CORR and VS with SIM
unchanged, confirming that reward-based optimization jointly enhances
all dimensions.

In SFT Phase~2, incorporating CKA further improves melody adherence,
as reflected in F0-CORR gains. For the perturbation ablation (w/o
Dist), we remove the temporal dropout applied to the melody latent
$\tilde{\mathbf{h}}$. This causes severe intelligibility degradation,
as the unperturbed latent retains residual semantic information that
the model exploits to bypass genuine lyric generation. Temporal
dropout eliminates this leakage, forcing reliance on the abstract
melodic contour and preserving prosodic structure while allowing free
generation of modified lyrics.

\vspace{-0.2cm}
\section{Conclusion}

\begin{table}[!t]
  \centering
  \caption{Ablation Study on LyricEditBench. Best results are
  \textbf{bold}, second best \underline{underlined}.}
  \vspace{-0.2cm}
  \label{tab:ablation}
  \begingroup
  \fontsize{8pt}{9.6pt}\selectfont
  \setlength{\tabcolsep}{3pt}
  \renewcommand{\arraystretch}{1}
  \begin{tabular}{@{}ll*{8}{c}@{}}
    \toprule
    \multirow{2}{*}{\textbf{Lg}} &
    \multirow{2}{*}{\textbf{Variant}} &
    \multicolumn{4}{c}{\textbf{Melody Control}} &
    \multicolumn{4}{c}{\textbf{Sing Edit}} \\
    \cmidrule(lr){3-6} \cmidrule(lr){7-10}
    & & P $\downarrow$ & S $\uparrow$ & F $\uparrow$ & V $\uparrow$
    & P $\downarrow$ & S $\uparrow$ & F $\uparrow$ & V $\uparrow$ \\
    \midrule
    \multirow{6}{*}{ZH}
    & TTS Pretrain       & 0.41 & 0.57 & 0.01 & 0.50 & 0.37 & 0.59 &
    0.06 & 0.49 \\
    & SFT Phase1         & \textbf{0.05} & \textbf{0.68} & 0.03 &
    1.55 & \textbf{0.05} & 0.73 & 0.31 & 1.53 \\
    & SFT Phase2         & 0.08 & 0.63 & \underline{0.92} &
    \underline{1.57} & 0.11 & \underline{0.75} & \underline{0.95} &
    \underline{1.62} \\
    & \quad -w/o CKA  & 0.08 & \underline{0.64} & 0.91 &
    \underline{1.57} & 0.12 & \underline{0.75} & 0.93 & 1.61 \\
    & \quad -w/o Dist & 0.45 & 0.63 & \textbf{0.94} & 1.42 & 0.46 &
    \textbf{0.79} & \underline{0.95} & 1.55 \\
    & Full Model          & \underline{0.06} & \underline{0.64} &
    \textbf{0.94} & \textbf{2.06} & \underline{0.08} &
    \underline{0.75} & \textbf{0.96} & \textbf{1.92} \\
    \midrule
    \multirow{6}{*}{EN}
    & TTS Pretrain       & 0.46 & 0.54 & 0.01 & 0.49 & 0.43 & 0.56 &
    0.00 & 0.50 \\
    & SFT Phase1         & \textbf{0.10} & \textbf{0.65} & 0.03 &
    1.07 & \textbf{0.11} & 0.73 & 0.40 & 1.13 \\
    & SFT Phase2         & \underline{0.14} & \underline{0.60} & 0.92
    & 1.17 & 0.19 & \underline{0.75} & \textbf{0.96} & \underline{1.21} \\
    & \quad -w/o CKA  & \underline{0.14} & \underline{0.60} & 0.90 &
    \underline{1.19} & 0.20 & \underline{0.75} & \underline{0.94} & 1.18 \\
    & \quad -w/o Dist & 0.48 & 0.58 & \textbf{0.95} & 1.00 & 0.49 &
    \textbf{0.78} & \textbf{0.96} & 0.98 \\
    & Full Model          & \textbf{0.10} & \underline{0.60} &
    \underline{0.93} & \textbf{1.52} & \underline{0.13} & 0.74 &
    \textbf{0.96} & \textbf{1.39} \\
    \bottomrule

  \end{tabular}
  \vspace{-0.5cm}
  \endgroup
\end{table}

We present YingMusic-Singer, a melody-controllable singing voice
editing model that synthesizes from a timbre reference, a
melody-providing singing clip, and modified lyrics without manual
alignment. Through curriculum training and GRPO-based reinforcement
learning, YingMusic-Singer achieves superior melody preservation and
lyric adherence on LyricEditBench, the first comprehensive benchmark
we introduce for this task, demonstrating strong potential for
practical end-to-end singing voice editing.

\section{Generative AI Use Disclosure}
Generative AI tools are used solely for linguistic refinement and
play no role in methodology, experimentation, interpretation, or the
production of scientific results. The authors bear full intellectual
responsibility for all content in this manuscript.

\bibliographystyle{IEEEtran}
\bibliography{mybib}

\end{document}